\documentclass[conference]{IEEEtran}

\usepackage{acronym}
\usepackage{amsfonts}
\usepackage{color, soul}
\usepackage{tikz}
\usetikzlibrary{arrows,positioning,shapes.geometric,decorations.pathreplacing}
\usepackage{multirow}
\usepackage{multicol}
\usepackage{mathtools}
\usepackage{algorithm, algpseudocode}
\usepackage{bm, dsfont}
\usepackage{subcaption}

\acrodef{bpsk}[BPSK]{binary phase-shift keying}
\acrodef{awgn}[AWGN]{additive white Gaussian noise}
\acrodef{sc}[SC]{successive-cancellation}
\acrodef{scl}[SCL]{SC list}
\acrodef{crc}[CRC]{cyclic redundancy check}
\acrodef{mdp}[MDP]{Markov decision process}
\acrodef{llr}[LLR]{log-likelihood ratio}
\acrodef{fer}[FER]{frame-error rate}
\acrodef{pm}[PM]{path metric}
\acrodef{rl}[RL]{reinforcement-learning}
\acrodef{td}[TD]{temporal difference}
\acrodef{snr}[SNR]{signal-to-noise ratio}
\acrodef{cscl}[CA-SCL]{CRC-aided SCL}

\newcommand{\PM}{\mathrm{PM}}
\newcommand{\sgn}{\mathrm{sgn}}

\makeatletter
\newcommand\fs@betterruled{%
  \def\@fs@cfont{\bfseries}\let\@fs@capt\floatc@ruled
  \def\@fs@pre{\vspace*{6pt}\hrule height.8pt depth0pt \kern2pt}%
  \def\@fs@post{\kern2pt\hrule\relax}%
  \def\@fs@mid{\kern2pt\hrule\kern2pt}%
  \let\@fs@iftopcapt\iftrue}
\floatstyle{betterruled}
\restylefloat{algorithm}
\makeatother
\begin{document}

\title{Construction of Polar Codes with\\Reinforcement Learning}

\author{\IEEEauthorblockN{Yun~Liao, Seyyed~Ali~Hashemi, John~Cioffi, Andrea~Goldsmith}
\IEEEauthorblockA{Department of Electrical Engineering, Stanford University, USA\\
yunliao@stanford.edu, ahashemi@stanford.edu, cioffi@stanford.edu, andrea@wsl.stanford.edu}
}

\maketitle

\begin{abstract}
This paper formulates the polar-code construction problem for the successive-cancellation list~(SCL) decoder as a maze-traversing game, which can be solved by reinforcement learning techniques. The proposed method provides a novel technique for polar-code construction that no longer depends on sorting and selecting bit-channels by reliability. Instead, this technique decides whether the input bits should be frozen in a purely sequential manner. The equivalence of optimizing the polar-code construction for the SCL decoder under this technique and maximizing the expected reward of traversing a maze is drawn. Simulation results show that the standard polar-code constructions that are designed for the successive-cancellation decoder are no longer optimal for the SCL decoder with respect to the frame error rate. In contrast, the simulations show that, with a reasonable amount of training, the game-based construction method finds code constructions that have lower frame-error rate for various code lengths and decoders compared to standard constructions.
\end{abstract}

\IEEEpeerreviewmaketitle

\section{Introduction}

Polar codes can achieve the capacity of any binary-input symmetric channel with low-complexity encoding and \ac{sc} decoding algorithms when the code length tends towards infinity \cite{arikan}. These codes were recently adopted in the control channel of the enhanced mobile broadband (eMBB) scenario for the fifth generation mobile-communications 5G standard, which requires codes with short block lengths \cite{3gpp_polar}. Since \ac{sc} decoding does not result in a satisfactory error-correction performance for short-block-length polar codes, \ac{sc} decoding variants, such as \mbox{\ac{scl}} decoding concatenated with a \mbox{\ac{crc} \cite{tal_list}}, are used to decode short-block-length polar codes.

\ac{sc} decoding and its variants are sequential-decoding algorithms that progress bit by bit. The polar-code encoding process divides the encoder input bits into two sets based on the underlying synthetic channels' reliability. One set of the input bits, corresponding to the more reliable synthetic channels, is assigned to carry information bits. The other set, corresponding to the less reliable synthetic channels, carries predefined values known to the decoder. The problem of finding the synthetic channels' reliability and dividing the input bits into two sets is called code construction. In fact, polar-code construction works to provide the best error-correction performance for a specific transmission channel with its associated specific sequential decoder.

Several techniques have been proposed to construct polar codes with \ac{sc} decoding. The Bhattacharyya parameter was first used in \cite{arikan} to construct polar codes. Density evolution \cite{mori1,mori2}, Gaussian approximation of density evolution \cite{trifonov_GA}, and upgrading/downgrading of channels \cite{tal_construction,pedarsani} were also used to construct polar codes with SC decoding. A universal partial order based upon the reliability of synthetic channels was found in \cite{bardet,schurch}, and it was shown in \cite{mondelli_complexity} that by using these universal partial orders, the complexity of polar-code construction is sublinear with the block length. Moreover, $\beta$-expansion was used in \cite{beta} to construct polar codes for different channels.

All the aforementioned polar-code-construction techniques are for use with \ac{sc} decoding. However, polar-code construction with \ac{sc} decoding does not necessarily result in the best error-correction performance under variants of \ac{sc} decoding such as \ac{scl} decoding \cite{hashemi_part}. To address this issue, heuristic methods such as Monte-Carlo simulations \cite{sun_MC,qin_MC} were used to construct polar codes for other decoding algorithms. Artificial-intelligence techniques have evolved as promising candidates to construct polar codes. In particular, genetic algorithms and machine learning were used to construct codes for specific decoders in \cite{elkelesh_GA,huang_AI}, respectively.

Different from most existing polar-code construction techniques that sort bit channels by reliability and then pick the most reliable ones, this paper explores the sequential-decoding process and approaches polar-code construction from a novel perspective. In particular, this paper proposes a technique whereby polar-code construction maps to a game, in which the agent is trained to traverse a maze. The connection between the maze-traversing game and the \ac{sc}-based decoding is detailed that minimizes the \ac{fer} by maximizing the game's expected return. The \ac{rl} algorithm SARSA$(\lambda)$ \cite{sutton1998introduction} is adopted in solving the game. Simulation results show that with a moderate amount of training, the game-based polar-code constructions can match current standard constructions for \ac{sc} decoding, and outperform the standard construction with \ac{scl} decoding. Moreover, we show that the \ac{fer} gap between the game-based constructions and the standard constructions increases with the list size in \ac{scl} decoding.
%Moreover, for longer codes, the game-based constructions' \ac{fer}s are lower than the standard constructions' \ac{fer}s with both \ac{sc} and \ac{scl} decoders.

The remainder of this paper is organized as follows: Section~\ref{sec:prel} reviews the preliminaries. Section~\ref{sec:game} details the proposed polar-code construction game. Section~\ref{sec:rl} explains the \ac{rl} algorithm that solves the game. Section~\ref{sec:simu} presents simulation results. Finally, Section~\ref{sec:conc} concludes the paper.

\section{Preliminaries}
\label{sec:prel}

\subsection{Polar Codes}
A polar code $\mathcal{P}(N, K)$ of length $N = 2^n$ is constructed by applying a linear transformation to input bit vector $\bm{u} = (u_0, u_1, \ldots, u_{N - 1})$ to obtain codeword $\bm{x} = \bm{u} \mathbf{G}^{\otimes n} = (x_0, x_1, \ldots, x_{N - 1})$, where $\mathbf{G}^{\otimes n}$ is the \mbox{$n$-th} Kronecker power of the polarizing kernel matrix $\mathbf{G} = \left[\begin{smallmatrix}
    1 & 0 \\
    1 & 1
\end{smallmatrix}\right]$. $\bm{x}$ is then modulated and transmitted through the channel. The input $\bm{u}$ contains a set $\mathcal{I}$ of $K$ nonfrozen bits, which need to be recovered, and a set $\mathcal{F}$ of $N - K$ frozen bits, whose positions and values are known to both the encoder and the decoder. The $K$ nonfrozen bits are divided into $A$ information bits and $P=K-A$ \ac{crc} bits. If no \ac{crc} is used, then $A = K$. The \emph{construction} of a polar code $\mathcal{P}(N, K)$ refers to the selection of $K$ nonfrozen bit positions.

This paper only considers \ac{bpsk} modulation for an \ac{awgn} channel. The values of the frozen bits are fixed as zero. %The algorithm can be easily extended to other modulation schemes and channel models.

\subsection{\ac{sc}-Based Decoding}
\begin{figure}
    \centering
    \begin{tikzpicture}[scale=2, thick]

  \fill[gray] (0,0) circle [radius=.05];

  \fill[gray] (-1,-.5) circle [radius=.05];
  \fill[gray] (1,-.5) circle [radius=.05];

  \fill[gray] (-1.5,-1) circle [radius=.05];
  \fill[gray] (-.5,-1) circle [radius=.05];
  \fill[gray] (.5,-1) circle [radius=.05];
  \fill[gray] (1.5,-1) circle [radius=.05];

  \draw (-1.75,-1.5) circle [radius=.05];
  \draw (-1.25,-1.5) circle [radius=.05];
  \draw (-.75,-1.5) circle [radius=.05];
  \fill (-.25,-1.5) circle [radius=.05];
  \draw (.25,-1.5) circle [radius=.05];
  \fill (.75,-1.5) circle [radius=.05];
  \fill (1.25,-1.5) circle [radius=.05];
  \fill (1.75,-1.5) circle [radius=.05];

  \draw (0,-.05) -- (-1,-.45);
  \draw (0,-.05) -- (1,-.45);

  \draw (-1,-.55) -- (-1.5,-.95);
  \draw (-1,-.55) -- (-.5,-.95);
  \draw (1,-.55) -- (.5,-.95);
  \draw (1,-.55) -- (1.5,-.95);

  \draw (-1.5,-1.05) -- (-1.75,-1.45);
  \draw (-1.5,-1.05) -- (-1.25,-1.45);
  \draw (-.5,-1.05) -- (-.75,-1.45);
  \draw (-.5,-1.05) -- (-.25,-1.45);
  \draw (.5,-1.05) -- (.25,-1.45);
  \draw (.5,-1.05) -- (.75,-1.45);
  \draw (1.5,-1.05) -- (1.25,-1.45);
  \draw (1.5,-1.05) -- (1.75,-1.45);

  \draw [very thin,gray,dashed] (-2,0) -- (2,0);
  \draw [very thin,gray,dashed] (-2,-.5) -- (2,-.5);
  \draw [very thin,gray,dashed] (-2,-1) -- (2,-1);
  \draw [very thin,gray,dashed] (-2,-1.5) -- (2,-1.5);

  \node at (-2.2,.3) {layer};
  \node at (-2.2,0) {3};
  \node at (-2.2,-.5) {2};
  \node at (-2.2,-1) {1};
  \node at (-2.2,-1.5) {0};

  \draw [->] (-.12,-.05) -- (-1,-.4) node [above=-.1cm,midway,rotate=25] {$\alpha$};
  \draw [->] (-.88,-.45) -- (0,-.1) node [below=-.1cm,midway,rotate=25] {$\beta$};

  \draw [->] (-1.06,-.55) -- (-1.5,-.9) node [above=-.1cm,midway,rotate=40] {$\alpha^{\rm left}$};
  \draw [->] (-1.44,-.95) -- (-1.0,-0.6) node [below=-.1cm,near start,rotate=40] {$\beta^{\rm left}$};

  \draw [<-] (-.94,-.55) -- (-.5,-.9) node [above=-.1cm,midway,rotate=-40] {$\beta^{\rm right}$};
  \draw [<-] (-.56,-.95) -- (-0.975,-.625) node [below=-.1cm,midway,rotate=-40] {$\alpha^{\rm right}$};

\end{tikzpicture}
    \caption{Binary tree representation for the \ac{sc} decoding of $\mathcal{P}(8,4)$ code.}
    \vspace{-1em}
    \label{fig:sc_binary}
\end{figure}
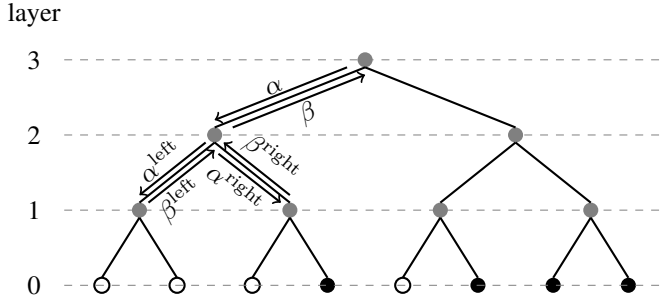

\ac{sc} decoding and its variants decode the $k$-th input bit based on the received signal $\bm{y}$ and the previously decoded bits $\hat{\bm{u}}^{k - 1} = (\hat{u}_0,\hat{u}_1,\ldots,\hat{u}_{k-1})$ via the conditional \ac{llr} value $\mbox{LLR}(u_k | \bm{y}, \hat{\bm{u}}^{k - 1})$ \cite{arikan}. Fig.~\ref{fig:sc_binary} illustrates the calculation of the conditional \ac{llr} as a binary tree search. Each node in layer $m$ corresponds to $2^m$ bits. The soft messages $\alpha$ that contain the \ac{llr} values are passed from a parent node to its child nodes, while the hard-bit estimates $\beta$ are passed upwards from a child node to its parent. The messages flow through a node in the following order: get $\alpha$ from its parent; send $\alpha^{\rm left}$ to its left child; get back $\beta^{\rm left}$; send $\alpha^{\rm right}$ to its right child; get back $\beta^{\rm right}$; and finally send $\beta$ back to its parent.

The $i$-th entry in $\alpha^{\rm left}, \alpha^{\rm right} \in \mathbb{R}^{2^{m - 1}}$ sent from a node in layer $m$ to its left and right children are calculated as, respectively,
\begin{equation}
    \alpha_i^{\rm left} = \sgn (\alpha_i \alpha_{i + 2^{m - 1}}) \cdot \min (|\alpha_i|, |\alpha_{i + 2^{m - 1}}|),
\end{equation}
\begin{equation}
    \alpha_i^{\rm right} = \alpha_{i + 2^{m - 1}} + (1 - 2\beta_i^{\rm left}) \alpha_i,
\end{equation}
and the initial $\alpha^{\rm left}$ sent from the root node contains the \ac{llr} values of the received $N$ symbols. The $i$-th entry in $\beta \in \{0,1\}^{2^m}$ to be returned to the node's parent is given by
\begin{equation}
    \beta_i = \begin{cases}
    \beta_i^{\rm left} \oplus \beta_i^{\rm right}, & \mbox{if } i < 2^{m - 1}, \\
    \beta_{i - 2^{m-1}}^{\rm right}, & \mbox{otherwise},
    \end{cases}
\end{equation}
where $\oplus$ denotes binary addition. 

When the $k$-th leaf node receives $\alpha$ from its parent, the decoder decodes the $k$-th bit $\hat{u}_k$ as
\begin{equation}\label{eq:sc_dec}
    \beta = \hat{u}_k = \begin{cases}
    0, & \mbox{if } k \in \mathcal{F} \mbox{ or } \alpha \ge 0, \\
    1, & \mbox{otherwise}.
    \end{cases}
\end{equation}

The \ac{scl} decoder improves the decoding performance of \ac{sc} decoder by keeping up to $L$ most likely decoding paths in parallel. Whenever a nonfrozen bit is encountered, both possible values, $0$ and $1$, are considered. A \ac{pm} is used to evaluate each decoding path. In particular, at the $k$-th leaf node, the \ac{pm} for the $l$-th path with $\hat{\bm{u}}^{k, (l)} = (\hat{u}_0^{(l)}, \ldots, \hat{u}_k^{(l)})$ is
\begin{equation}
    \PM_k^{(l)} = \begin{cases}
    \PM_{k - 1}^{(l)}, & \mbox{if } \hat{u}_k^{(l)} = \frac{1 - \sgn(\alpha^{(l)})}{2}, \\
    \PM_{k - 1}^{(l)} + |\alpha^{(l)}|, & \mbox{otherwise},
    \end{cases}
\end{equation}
where $\alpha^{(l)}$ is the soft message passed to the leaf node though the $l$-th path. After computing the \ac{pm}s for all possible paths, the $L$ paths with the lowest \ac{pm}s survive, and the others are dropped. 

When the decoding process ends, one codeword needs to be selected from the list. There are several ways to select the final decoder result:
\begin{enumerate}
    \item {\bf Pure \ac{scl}}: the decoder selects the codeword with the smallest \ac{pm} from the list.
    \item {\bf\ac{cscl}}: the decoder decodes to the codeword with the smallest \ac{pm} among the candidates that pass the \ac{crc} check.
    \item {\bf \ac{scl}-Genie}: the decoder decodes to the correct codeword as long as it is in the list.
\end{enumerate}

\begin{figure}[t!]
    \centering
    \includegraphics[width=.85\linewidth]{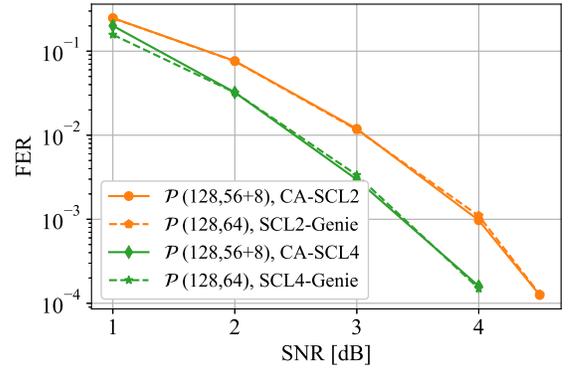}
    %\caption{\ac{fer} comparison between \ac{scl}-Genie decoding of $\mathcal{P}(128,64)$ and \ac{cscl} decoding of $\mathcal{P}(128, 56+8)$. \ac{cscl}$L$ denotes the \ac{cscl} decoder with a list size $L$, and \ac{scl}$L$-Genie denotes the \ac{scl}-Genie decoder with a list size $L$.}
    \caption{\ac{fer} comparison between \ac{scl}-Genie decoding of $\mathcal{P}(128,64)$ and \ac{cscl} decoding of $\mathcal{P}(128, 56+8)$. \ac{scl}$L$ denotes the \ac{scl}-based decoder with a list size $L$.}
    \vspace{-1em}
    \label{fig:genie}
\end{figure}

Although the \ac{scl}-Genie decoder cannot be implemented in practice, it is adopted during training where the correct codewords are known for the training samples. The main reasons are as follows. First, for a given construction, the \ac{cscl} decoder for $\mathcal{P}(N,A+P)$ yields almost identical decoding performance as the \ac{scl}-Genie decoder for $\mathcal{P}(N,K)$ with $K = A+P$, as long as $P$ is moderately large. This is verified in Fig.~\ref{fig:genie}, which shows the performance of the \ac{cscl} decoder for $\mathcal{P}(128, 56+8)$ and the corresponding performance of the \ac{scl}-Genie decoder for $\mathcal{P}(128, 64)$. Besides, the design of a good \ac{crc} given the construction of $\mathcal{P}(N, K)$ is a separate problem, which is beyond the scope of this work.

\section{Viewing Polar Code Construction as a Game}
\label{sec:game}

\subsection{Polar Code Construction Game}

The construction of $\mathcal{P}(N,K)$ is the selection of $K$ of $N$ nonfrozen bit positions. The selection procedure is equivalent to a maze-traversing game as follows:

\begin{itemize}
    \item \textbf{Environment}: A maze with height $N - K + 1$ and width $K + 1$. The states of the environment are defined as the cells $({\rm row}, {\rm col})$. The upper left cell indexed by $(0,0)$ is set as the start cell. The bottom right cell indexed by $(K, N - K)$ is set as the terminal cell. 
    \item \textbf{Rule}: Each game starts from the start cell. At each step, the agent takes an action $a$ that is either ``move \emph{down}'' $(a = 0)$ or ``move \emph{right}'' $(a = 1)$, and it is not allowed to depart the maze. The game ends when the agent reaches the terminal cell or when it receives a nonzero reward from the environment.
    \item \textbf{Reward}: The reward is associated with the \ac{scl}-Genie decoding process, which is further explained in Section~\ref{subsection:reward}.
    \item \textbf{Goal}: The agent attempts to find the best path that yields the highest expected return throughout the game.
\end{itemize}

\begin{figure}[t!]
    \centering
    \includegraphics[width=.54\linewidth]{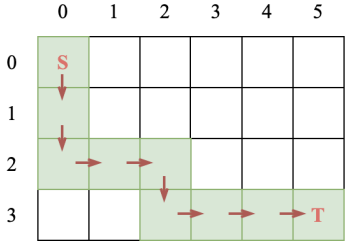}
    \vspace{-.5em}
    \caption{An example of a polar code construction game for $\mathcal{P}(8,5)$. The action list is $\{\text{down}, \text{down}, \text{right}, \text{right}, \text{down}, $ $\text{right}, \text{right}, \text{right}\}$. The corresponding nonfrozen bit positions are $\{2, 3, 5, 6, 7\}$.}
    \vspace{-1.5em}
    \label{fig:polar_maze}
\end{figure}

Each possible path that the agent can choose in the maze corresponds to a possible construction of $\mathcal{P}(N,K)$. In particular, the $k$-th bit is set as a frozen bit if the agent chooses the \emph{down} action at the $k$-th step; and it is set as an information bit if the agent chooses the \emph{right} action instead. Fig.~\ref{fig:polar_maze} is an example of the maze associated with the construction of $\mathcal{P}(8,5)$. Both the bit positions in the polar code and the steps in the game are indexed from $0$ to $N - 1$. 

\subsection{Reward Generation via \ac{scl}-Genie Decoding}
\label{subsection:reward}
The design of the instant reward at each step needs to satisfy the following requirements: (1) the reward needs to reflect how good the current action is in the short term; (2) the path with high expected return at the game's end should reflect a good polar-code construction, i.e., a construction that gives low \ac{fer}.

In the \ac{scl}-Genie decoding process, the bits are decoded sequentially. In particular, the decoding of the $k$-th bit is independent of how the frozen and nonfrozen bits are distributed after it, and given the \ac{pm}s and the survival paths at the $(k-1)$-th bit, the evolution of the survival paths and the \ac{pm}s at the $k$-th bit is independent of the distribution of frozen and nonfrozen positions other than the $k$-th bit itself. The sequential nature of the \ac{scl}-Genie decoder suggests that each step's instant reward can be set along with the decoding process. The detailed reward generation process is given in Algorithm~\ref{alg:reward}. 

In the reward-generation process, the transmitted codeword is fixed to the all-zero codeword because it is the only valid codeword for all possible polar-code constructions. Algorithm~\ref{alg:reward} shows that the selected actions are penalized when the \ac{scl}-Genie decoder fails to decode the correct codeword, i.e., when the \ac{scl}-Genie decoder drops the correct codeword from the list during the decoding process. As such, by selecting a maze path with a high expected return, or equivalently, a small expected penalty, the agent implicitly chooses a polar-code construction with low expected \ac{fer}. The \ac{sc}-based decoders only append new bits after the already decoded bit stream, and no previous decisions will be altered. Therefore, once the correct codeword is dropped from the list at step $k$, a frame error must occur, and the actions after step $k$ cannot repair that result. In other words, it is unreasonable to prefer any actions over others after dropping the correct codeword at step $k$. Therefore, the game is designed to be terminated after the decoder drops the correct codeword, or equivalently, after the agent receives nonzero reward.

\begin{algorithm}[t]
    \caption{Reward generation at step $k$\label{alg:reward}}
    \hspace*{\algorithmicindent} \textbf{Input:} step $k$, action $a$ \\
    \hspace*{\algorithmicindent} \textbf{Global Variable:} PM list, survival path list \\
    \hspace*{\algorithmicindent} \textbf{Output:} reward $r$, termination flag $F$
    \begin{algorithmic}[1]
        \If {$k = 0$} 
            \State Transmit all-zero codeword through the channel
            \State Initialize the \ac{scl}-Genie decoder with the received \ac{llr}
        \EndIf
        \If {$a = 0$}
            \State Decode the $k$-th bit as if it is a frozen bit
            \State Update the \ac{pm} list and the survival paths
        \Else
            \State Decode the $k$-th bit as if it is a nonfrozen bit
            \State Update the \ac{pm} list and the survival paths
            \State Check if the all-zero codeword survives
            \If {all-zero codeword is dropped from the list}
                \State Set $r = -1$, $F = \mathrm{True}$
                \State Clear PM list, survival path list
                \State \Return $(r, F)$
            \EndIf
        \EndIf
        \State Set $r = 0$
        \If {$k = N-1$}
            \State Set $F = \mathrm{True}$
            \State Clear PM list, survival path list
        \EndIf
        \State Set $F = \mathrm{False}$
        \State \Return $(r, F)$
    \end{algorithmic}
\end{algorithm}

\section{Reinforcement Learning Algorithm}\label{sec:rl}
Here, a \ac{rl} technique solves the polar-code construction problem. This section introduces tabular \ac{rl} systems and then describes the SARSA$(\lambda)$ algorithm for agent training.

\subsection{Reinforcement Learning Basics}
A typical tabular \ac{rl} system contains an environment and an agent. At a given time $t$, the environment is at state $s_t \in \mathcal{S}$, and the agent takes an \emph{action} $a_t \in \mathcal{A}$ according to a \emph{policy} $\pi: \mathcal{S} \to \mathcal{A}$ to interact with the environment. The environment, stimulated by the agent's action, changes its state to $s_{t+1}$ and provides \emph{reward} $r_{t+1}$ to the agent. The agent accumulates the rewards as this interaction continues. The \emph{return} $R = \sum_{t = 0}^{T} \gamma^{t} r_{t+1}$ is the accumulated reward that the agent receives throughout the game, in which $\gamma \in (0,1]$ is the discount rate that describes how much the agent weights the future reward, and $T$ denotes the game's termination time. The agent's goal is to optimize its policy $\pi$ to maximize the expected return $\mathbb{E}[R]$. 

The agent's policy $\pi$ is commonly derived from a value function $Q: \mathcal{S} \times \mathcal{A} \to \mathbb{R}$, which approximates the expected return when taking action $a$ from state $s$ and then following policy $\pi$. The value function of any state-action pair $(s,a) \in \mathcal{S} \times \mathcal{A}$ under policy $\pi$ is defined as
\begin{equation}
    Q^{\pi}(s, a) = \mathbb{E}_{\pi}\left[\left.\sum_{\tau = 0}^{T-t-1}\gamma^{\tau} r_{t+\tau+1} \right\vert s_t = s, a_t = a \right],
\end{equation}
and it satisfies the dynamics
\begin{equation}
    Q^{\pi}(s,a) = \mathbb{E}\left[r_{t+1} + \gamma Q^{\pi}(s_{t+1}, \pi (s_{t+1}))|s_t = s, a_t = a\right].
\end{equation}

An $\epsilon$-greedy policy according to a value function $Q$ is defined as
\begin{equation}
    \pi(s) = \begin{cases}
    \displaystyle\arg \max_{a \in \mathcal{A}} Q(s, a), & \mbox{ w.p. } 1 - \epsilon, \\
    \mbox{random } a \in \mathcal{A}, & \mbox{ w.p. } \epsilon.
    \end{cases}
\end{equation}
\iffalse
The key to maximizing the expected return is the value function $Q: \mathcal{S} \times \mathcal{A} \to \mathbb{R}$, which approximates the expected return when taking action $a$ from state $s$ and then following policy $\pi$. The value function satisfies
\begin{equation}
    Q(s_t, a_t) = r_{t+1} + \gamma Q(s_{t+1}, a_{t+1}).
\end{equation}
With an accurate estimate of the value function, the optimal policy is
\begin{equation}
    a_t^* = \arg\max_{a \in \mathcal{A}} Q(s_t, a).
\end{equation}
\fi
\subsection{SARSA$(\lambda)$ with Eligibility Trace}
To learn a good policy, the agent updates the value function according to the rewards it receives from the environment. In this work, the agent uses the SARSA$(\lambda)$ algorithm with eligibility trace \cite{sutton1998introduction} to update the value function. 

An eligibility trace captures the current game's historical trace and assigns credit to every state-action pair. In particular, the eligibility trace initializes as
\begin{equation}
    E_0(s,a) = 0,~\forall s \in \mathcal{S}, a \in \mathcal{A},
\end{equation}
and it evolves as
\begin{equation}
    E_t(s,a) = \gamma \lambda E_{t - 1} (s,a) + \mathds{1} (s_t = s, a_t = a),
\end{equation}
where $\mathds{1}(\cdot)$ is the indicator function, and the parameter $\lambda \in [0,1]$ indicates how much an agent would change the value function of an early state in the game according to a reward that is received later. A large $\lambda$ means that the agent traces back deeply and updates the value function of historical states at each step. A small $\lambda$ means that at each step, only the values of several recent states will be changed with the newly received reward.

During training, the agent maintains a table of value functions $Q$, and uses the $\epsilon$-greedy policy, where the exploration rate $\epsilon$ decreases over training episodes. At time $t$ in one episode, the agent is at state $s_t$ and takes action $a_t$ according to the $\epsilon$-greedy policy based on the current value functions. Upon receiving the reward $r_{t + 1}$, the \ac{td} error is defined as
\begin{equation}
    \delta_t = r_{t + 1} + \gamma Q(s_{t+1}, a') - Q(s_t, a_t),
\end{equation}
in which $s_{t+1}$ is the next state after taking action $a_t$ at time $t$, and action $a'$ is selected according to the agent's current \mbox{$\epsilon$-greedy} policy from state $s_{t+1}$. The \ac{td} error roughly indicates how much the estimated value function deviates from the real reward. The value function is then updated as
\begin{equation}
    Q(s, a) \leftarrow Q(s,a) + \rho \delta_t E_t(s, a),~\forall s,a,
\end{equation}
where $\rho$ is the learning rate. The agent then selects and takes action $a_{t+1}$ according to the $\epsilon$-greedy policy based on the updated value functions from state $s_{t + 1}$, and proceeds to the next step.

\subsection{Equivalence to Polar Code Construction Problem}
The value function of the state-action pair $(s,a)$, by definition, is the agent's expected return after taking action $a$ at state $s$. The reward generating process described in Section~\ref{subsection:reward} indicates that the expected return after $(s,a)$ is
\begin{equation}\nonumber
    \begin{split}
        \mathbb{E}[R] &= 0 \times \Pr(\mbox{correct codeword survives}) + \\
        &(-1) \times \Pr(\mbox {correct codeword dropped afterwards}) \\
        &= - \Pr(\mbox {correct codeword dropped afterwards}) \\
        &= - \Pr(\mbox{frame error} | \mbox{correct decoding up to state }s).
    \end{split}
\end{equation}
Therefore, by learning the strategy that maximizes the expected return at the start state $s=(0,0)$, the agent is in fact learning to construct the polar code in the optimal way that minimizes the \ac{fer} under the \ac{scl}-Genie decoder. 

Recall that as shown in Fig.~\ref{fig:genie}, the \ac{fer} of $\mathcal{P}(N, K)$ under the \ac{scl}-Genie decoder is almost identical to the \ac{fer} of $\mathcal{P}(N, A+P)$ under the \ac{cscl} decoder when $K=A+P$. Therefore, the learned code construction from the game is nearly optimal for $\mathcal{P}(N, A+P)$ under the \ac{cscl} decoder.

\iffalse
\begin{figure}[t!]
    \centering
    \includegraphics[width=.8\linewidth]{Figures/value_fun.eps}
    \caption{Value function along the selected path for the $\mathcal{P}(16, 8)$ code construction.}
    \label{fig:value_fun}
\end{figure}
\fi

\section{Simulation Results}
\label{sec:simu}

To illustrate the performance of the proposed game-based polar-code construction method, the game's learned code constructions are evaluated under either the pure \ac{scl} decoder or the \ac{cscl} decoder. For each evaluation case, the number of simulated transmissions is such that the number of observed frame errors is at least $500$. The \ac{fer} performance of the learned code constructions is compared to the constructions given by the method in \cite{tal_construction}. In particular, since the polar-code constructions depend on the channel condition, the reported \ac{fer} performance uses the code construction that is either designed (for the method in \cite{tal_construction}) or trained (for the game-based method) at the given \ac{snr} level.

The parameters of the SARSA$(\lambda)$ algorithm are selected as the following. The discount rate $\gamma = 1$ since the agent cares about the \ac{fer} in the end, and having the \ac{scl} decoder drop the correct codeword at any step matters the same to the agent. The eligibility decay factor is $\lambda = 0.3$ for constructing the $\mathcal{P}(16,8)$ code, and $\lambda = 0.75$ for constructing the $\mathcal{P}(128,56+8)$ code.
\iffalse
$\lambda = 0$ for the \ac{sc} decoder; $\lambda = 0.3$ for the \ac{scl} and \ac{cscl} decoders with a list size of $2$; and $\lambda = 0.5$ for the \ac{scl} and \ac{cscl} decoders with a list size of $4$. The parameter $\lambda$ increases with the list size because it indicates how much the agent penalizes the historical actions for the current reward or penalty. For the \ac{sc} decoder, dropping the correct codeword at step $k$ is independent of how the previous $k-1$ actions are selected. For the \ac{scl} decoder, however, dropping the correct codeword at step $k$ indicates that there are already many decoding paths that are evolving in parallel, and some of them are competitive with the correct one. Therefore, the previous actions need to be penalized for this situation.
\fi

\subsection{\ac{fer} performance}
\begin{figure}[t!]
    \centering
    \includegraphics[width=.8\linewidth]{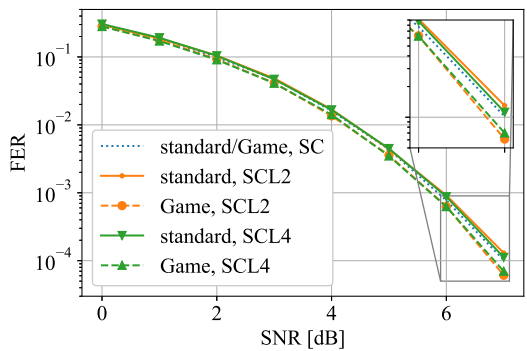}
    %\vspace{-.5em}
    \caption{Comparison of \ac{fer} for $\mathcal{P}(16,8)$ code under \ac{sc} and pure \ac{scl} decoder.}
    %\vspace{-1em}
    \label{fig:16_8_code_pure}
\end{figure}%
\begin{figure}[t!]
    \centering
    \begin{subfigure}{.49\linewidth}
        \centering
        \includegraphics[width=\linewidth]{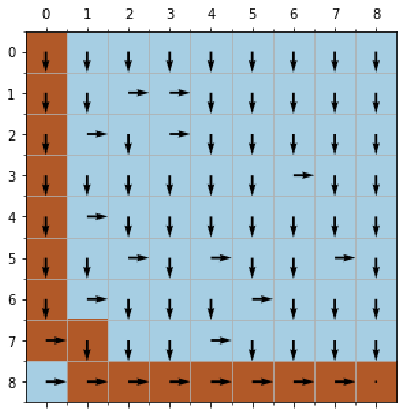}
        \caption{}
        \label{fig:16_8_1_maze}
    \end{subfigure}
    \begin{subfigure}{.49\linewidth}
        \centering
        \includegraphics[width=\linewidth]{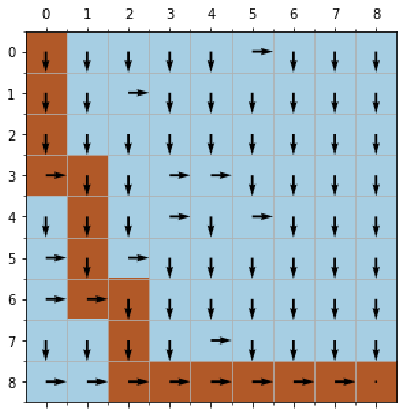}
        \caption{}
        \label{fig:16_8_2_maze}
    \end{subfigure}
    \caption{Path selected for $\mathcal{P}(16, 8)$ code at ${\rm SNR}=0$ dB for (a) \ac{sc} decoder; (b) \ac{scl} decoder with a list size of 2.}
    \vspace{-1em}
\end{figure}

Fig.~\ref{fig:16_8_code_pure} shows the \ac{fer} performance of the learned $\mathcal{P}(16,8)$ code compared to the standard construction given in \cite{tal_construction} under the \ac{sc} decoder as well as the pure \ac{scl} decoders with a list size $2$ and $4$. Under \ac{sc} decoder, the game-based constructions at each evaluated \ac{snr} level match the construction given by the method in \cite{tal_construction}. Fig.~\ref{fig:16_8_1_maze} shows the learned policy of the polar construction game under \ac{sc} decoder when ${\rm SNR} = 0$~dB. In particular, the arrow in each cell shows the best action when the agent is in that cell. The highlighted cells are the selected path from the start cell $(0,0)$ to the terminate cell $(8,8)$. It can be seen from the figure that the action ``move right'' is selected at steps $\{7,9,10,11,12,13,14,15\}$, which are known to be the most reliable $8$ bit-channels for a polar code with codeword length of $16$.

When the \ac{scl} decoders are used, the game-based constructions do not match the constructions of \cite{tal_construction}. Fig.~\ref{fig:16_8_2_maze} shows the selected path under \ac{scl}-Genie decoder with a list size $2$ at ${\rm SNR} = 0$~dB. The corresponding selected nonfrozen positions in this case are $\{3,7,10,11,12,13,14,15\}$. The \ac{fer} of the learned code constructions under pure \ac{scl} decoders without \ac{crc} is slightly better than the method in \cite{tal_construction} at every evaluated \ac{snr} level. The observation that the agent learns better code constructions than the standard construction under \ac{scl} decoders verifies that ranking and picking the most reliable $K$ bit-channels is no longer optimal for \ac{scl} decoders.

\begin{figure}[t!]
    \centering
    \includegraphics[width=\linewidth]{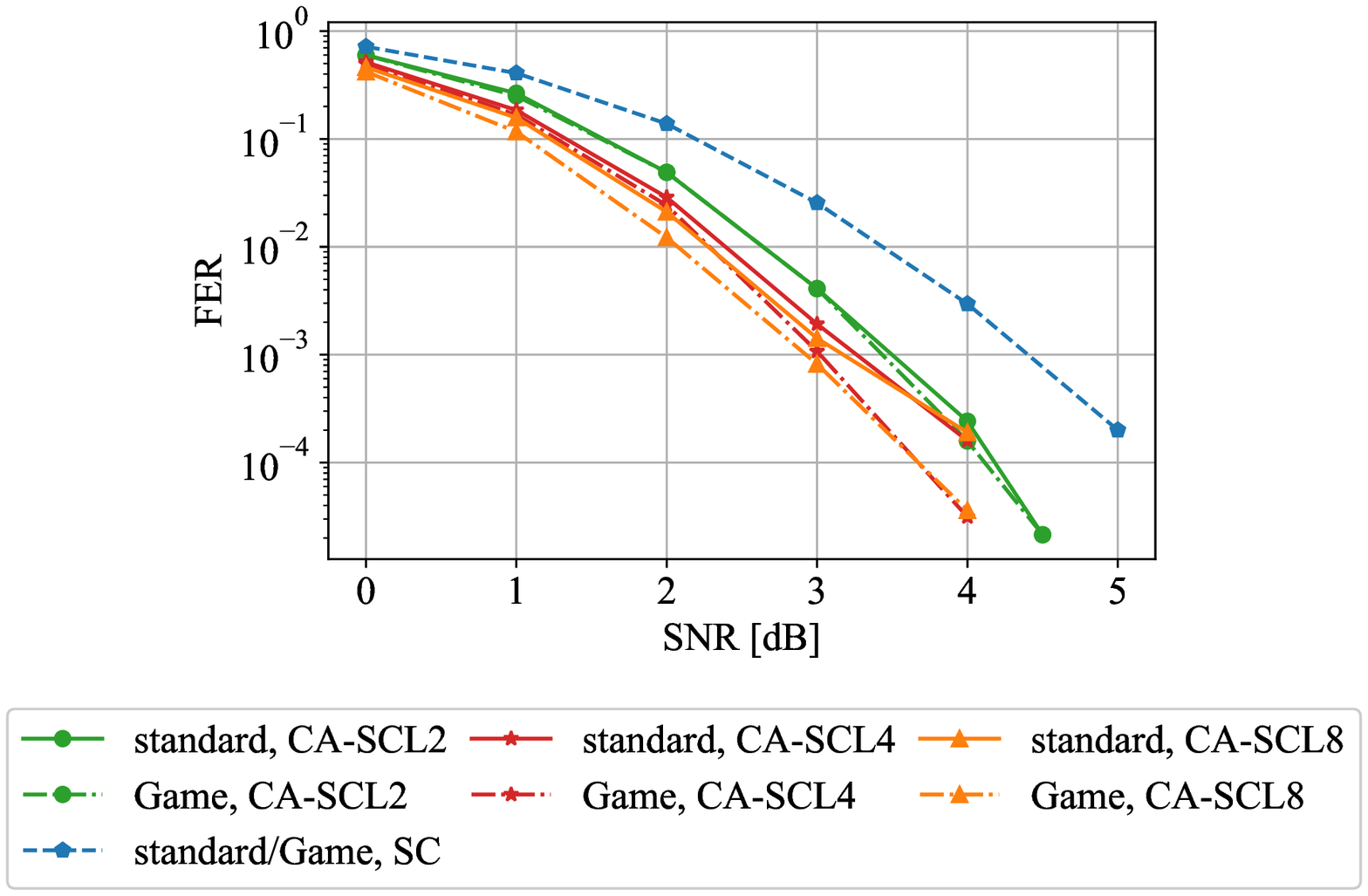}
    \caption{\ac{fer} performance comparison of $\mathcal{P}(128, 60+4)$ with \ac{scl} decoding.}
    \label{fig:128_56_code}
    \vspace{-1em}
\end{figure}%
\iffalse
\begin{figure}[t!]
    \centering
    \includegraphics[width=.8\linewidth]{Figures/128_56_constructions.eps}
    \caption{Learned polar-code constructions of $\mathcal{P}(128, 56+8)$ code at ${\rm SNR} = 4$~dB. On $y$-axis, $\mathcal{I}$ denotes nonfrozen positions, and $\mathcal{F}$ denotes frozen positions.}
    \label{fig:128_56_construction}
\end{figure}
\fi
The proposed game-based construction's advantage becomes more apparent for long codewords and when the \ac{crc} is used together with the selection of nonfrozen bit positions. Fig.~\ref{fig:128_56_code} illustrates the performance of the learned $\mathcal{P}(128, 60+4)$ codes. The game-based constructions match the standard constructions in \cite{tal_construction} under \ac{sc} decoding. This is expected since the standard constructions in \cite{tal_construction} are optimized for SC decoding. For the \ac{scl} decoder with list size $2$, the game-based constructions achieve almost the same performance as the standard ones. However, when the list size of \ac{scl} decoders is increased to $4$ and $8$,
%Fig.~\ref{fig:128_56_code} shows that the game-based algorithm finds better constructions than the standard construction in \cite{tal_construction}, even under the \ac{sc} decoder, for which the constructions in \cite{tal_construction} are designed. The
the game-based constructions outperform the ones in \cite{tal_construction} over the entire range of evaluated \ac{snr}. When the \ac{crc} protects the correct codeword in the \ac{scl} decoder's final candidate list, the game-based construction method shows a clear gain over the standard construction method. This is expected because the game optimizes the probability of keeping the correct codeword in the decoding list. A reliable \ac{crc} ensures that the decoder actually finds the correct codeword with high probability as long as it is in the final list. 

\iffalse
\textcolor{blue}{Fig.~\ref{fig:128_56_construction} compares the learned polar-code constructions of the $\mathcal{P}(128, 56+8)$ code for the \ac{sc} decoder and the \ac{scl} with a list size 2 at ${\rm SNR} = 4$ dB with the standard construction in \cite{tal_construction} at the same \ac{snr} level. The learned constructions }\hl{Need to talk about the new figure.}
\fi

\subsection{Efficiency of the Game-Based Construction Method}
Unlike the conventional construction methods that need to evaluate and rank the bit channels, the proposed game-based construction method selects the combination of nonfrozen bit positions altogether, without explicitly estimating the bit error rate on each bit-channel. The advantage of selecting the combination together in the polar-code-construction game is twofold: (1) as shown in the \ac{fer} comparisons, selecting the nonfrozen bit positions according to their ranking is not the optimal method under \ac{scl} decoders; (2) it avoids the need of running large-volume Monte-Carlo simulations on every bit-channel, which is so far the most accurate way to get a reliable bit-channel ranking, especially with long codewords. These Monte-Carlo simulations can be prohibitively expensive because the bit error rates on the bit-channels can be very small and close to each other.

The SARSA$(\lambda)$ algorithm updates the value function at every state along the selected path for each training sample fed into the system. By doing so, about $N$ value functions update with a single pass of one training sample. Most of the suboptimal paths are eliminated quickly at the beginning of the training, and the rest of the training distinguishes between several candidate paths that yield similar expected returns. The training process is highly efficient in terms of the number of training samples. To construct $\mathcal{P}(16, 8)$, only $2000$ samples are used during training, and each training sample is only used once. To construct $\mathcal{P}(128, 60+4)$ that are evaluated in Fig.~\ref{fig:128_56_code}, the training needs no more than $200,000$ training samples, with only one pass of each sample, and it is observed that the training usually converges before feeding $80,000$ training samples into the system. As a comparison, the methods described in \cite{huang_AI} converge in about $10,000$ iterations but needs to run Monte-Carlo simulations to estimate the \ac{fer} at each iteration, which needs at least $1/{\rm FER}$ samples. Therefore, the proposed game-based method is highly efficient compared to the methods in \cite{huang_AI} in terms of the number of training samples.

%\subection{}
\section{Summary}
\label{sec:conc}
This paper formulated the polar-code construction problem for the \ac{scl} decoder as a maze-traversing game, in which the game's expected return indicates the \ac{fer} of the selected polar-code construction. The tabular \ac{rl} algorithm SARSA$(\lambda)$ was adopted to solve the game. The inherent equivalence of the polar code construction problem and the game was revealed. Simulation results showed that the game-based constructions matched the standard polar code constructions under the \ac{sc} decoder for short codes. For short codes under the \ac{scl} decoder and longer codes under both the \ac{sc} and \ac{scl} decoders, the game-based method was able to find polar code constructions that outperform the standard constructions significantly. Moreover, the game-based method is very efficient during training in terms of the number of required training samples. Future work includes (1) evaluating the effect of channel mismatch during training and evaluation; (2) using neural networks or other deep learning techniques to estimate the value of the value function to improve the memory efficiency for the construction of longer codes; and (3) incorporating \ac{crc} verification in the reward process during training to further improve the constructions.

\section*{Acknowledgments}

This work is supported in part by ONR grant N00014-18-1-2191. S.~A.~Hashemi is supported by a Postdoctoral Fellowship from the Natural Sciences and Engineering Research Council of Canada (NSERC) and by Huawei.

%\vspace{-.2em}
\bibliographystyle{IEEEtran}
\bibliography{references}

\begin{thebibliography}{10}
\providecommand{\url}[1]{#1}
\csname url@samestyle\endcsname
\providecommand{\newblock}{\relax}
\providecommand{\bibinfo}[2]{#2}
\providecommand{\BIBentrySTDinterwordspacing}{\spaceskip=0pt\relax}
\providecommand{\BIBentryALTinterwordstretchfactor}{4}
\providecommand{\BIBentryALTinterwordspacing}{\spaceskip=\fontdimen2\font plus
\BIBentryALTinterwordstretchfactor\fontdimen3\font minus
  \fontdimen4\font\relax}
\providecommand{\BIBforeignlanguage}[2]{{%
\expandafter\ifx\csname l@#1\endcsname\relax
\typeout{** WARNING: IEEEtran.bst: No hyphenation pattern has been}%
\typeout{** loaded for the language `#1'. Using the pattern for}%
\typeout{** the default language instead.}%
\else
\language=\csname l@#1\endcsname
\fi
#2}}
\providecommand{\BIBdecl}{\relax}
\BIBdecl

\bibitem{arikan}
E.~{Arikan}, ``Channel polarization: A method for constructing
  capacity-achieving codes for symmetric binary-input memoryless channels,''
  \emph{IEEE Transactions on Information Theory}, vol.~55, no.~7, pp.
  3051--3073, 2009.

\bibitem{3gpp_polar}
3GPP, ``Final report of {3GPP TSG RAN WG1} \#87 v1.0.0,'' {R}eno, USA, Nov.
  2016.

\bibitem{tal_list}
I.~{Tal} and A.~{Vardy}, ``List decoding of polar codes,'' \emph{IEEE
  Transactions on Information Theory}, vol.~61, no.~5, pp. 2213--2226, 2015.

\bibitem{mori1}
R.~{Mori} and T.~{Tanaka}, ``Performance and construction of polar codes on
  symmetric binary-input memoryless channels,'' in \emph{2009 IEEE
  International Symposium on Information Theory}, 2009, pp. 1496--1500.

\bibitem{mori2}
------, ``Performance of polar codes with the construction using density
  evolution,'' \emph{IEEE Communications Letters}, vol.~13, no.~7, pp.
  519--521, 2009.

\bibitem{trifonov_GA}
P.~{Trifonov}, ``Efficient design and decoding of polar codes,'' \emph{IEEE
  Transactions on Communications}, vol.~60, no.~11, pp. 3221--3227, 2012.

\bibitem{tal_construction}
I.~{Tal} and A.~{Vardy}, ``How to construct polar codes,'' \emph{IEEE
  Transactions on Information Theory}, vol.~59, no.~10, pp. 6562--6582, 2013.

\bibitem{pedarsani}
R.~{Pedarsani}, S.~H. {Hassani}, I.~{Tal}, and E.~{Telatar}, ``On the
  construction of polar codes,'' in \emph{2011 IEEE International Symposium on
  Information Theory Proceedings}, 2011, pp. 11--15.

\bibitem{bardet}
M.~{Bardet}, V.~{Dragoi}, A.~{Otmani}, and J.~{Tillich}, ``Algebraic properties
  of polar codes from a new polynomial formalism,'' in \emph{2016 IEEE
  International Symposium on Information Theory (ISIT)}, 2016, pp. 230--234.

\bibitem{schurch}
C.~{Schürch}, ``A partial order for the synthesized channels of a polar
  code,'' in \emph{2016 IEEE International Symposium on Information Theory
  (ISIT)}, 2016, pp. 220--224.

\bibitem{mondelli_complexity}
M.~{Mondelli}, S.~H. {Hassani}, and R.~L. {Urbanke}, ``Construction of polar
  codes with sublinear complexity,'' \emph{IEEE Transactions on Information
  Theory}, vol.~65, no.~5, pp. 2782--2791, 2019.

\bibitem{beta}
G.~{He}, J.~{Belfiore}, I.~{Land}, G.~{Yang}, X.~{Liu}, Y.~{Chen}, R.~{Li},
  J.~{Wang}, Y.~{Ge}, R.~{Zhang}, and W.~{Tong}, ``Beta-expansion: A
  theoretical framework for fast and recursive construction of polar codes,''
  in \emph{GLOBECOM 2017 - 2017 IEEE Global Communications Conference}, 2017,
  pp. 1--6.

\bibitem{hashemi_part}
S.~A. {Hashemi}, M.~{Mondelli}, S.~H. {Hassani}, C.~{Condo}, R.~L. {Urbanke},
  and W.~J. {Gross}, ``Decoder partitioning: Towards practical list decoding of
  polar codes,'' \emph{IEEE Transactions on Communications}, vol.~66, no.~9,
  pp. 3749--3759, 2018.

\bibitem{sun_MC}
S.~{Sun} and Z.~{Zhang}, ``Designing practical polar codes using
  simulation-based bit selection,'' \emph{IEEE Journal on Emerging and Selected
  Topics in Circuits and Systems}, vol.~7, no.~4, pp. 594--603, 2017.

\bibitem{qin_MC}
M.~{Qin}, J.~{Guo}, A.~{Bhatia}, A.~{Guillén i Fàbregas}, and P.~H. {Siegel},
  ``Polar code constructions based on {LLR} evolution,'' \emph{IEEE
  Communications Letters}, vol.~21, no.~6, pp. 1221--1224, 2017.

\bibitem{elkelesh_GA}
A.~{Elkelesh}, M.~{Ebada}, S.~{Cammerer}, and S.~{ten Brink},
  ``Decoder-tailored polar code design using the genetic algorithm,''
  \emph{IEEE Transactions on Communications}, vol.~67, no.~7, pp. 4521--4534,
  2019.

\bibitem{huang_AI}
L.~{Huang}, H.~{Zhang}, R.~{Li}, Y.~{Ge}, and J.~{Wang}, ``{AI} coding:
  Learning to construct error correction codes,'' \emph{IEEE Transactions on
  Communications}, vol.~68, no.~1, pp. 26--39, 2020.

\bibitem{sutton1998introduction}
R.~S. Sutton, A.~G. Barto \emph{et~al.}, \emph{Introduction to reinforcement
  learning}.\hskip 1em plus 0.5em minus 0.4em\relax MIT press Cambridge, 1998,
  vol. 135.

\end{thebibliography}

\end{document}